\documentclass[12pt,twoside]{article}
\usepackage{epsfig}
\usepackage{graphicx}
\usepackage{amssymb}
\usepackage{url}
\usepackage{amsfonts,amsbsy}
\usepackage{amsmath}

\title{Electromagnetic Media with no Dispersion Equation}
\author{I.V. Lindell* and A. Favaro**} 
\date{*Department of Radio Science and Engineering\\ Aalto University, School of Electrical Engineering\\Espoo, Finland\\ 
** Institute of Theoretical Physics\\ University of Cologne, Germany\\
{\tt ismo.lindell@aalto.fi\\ favaro@thp.uni-koeln.de}}
\parskip=\medskipamount
\def\e{\begin{equation}} 
\def\f{\end{equation}} 
\def\ea{\begin{eqnarray}} 
\def\fa{\end{eqnarray}} 

\def\##1{{\mbox{\textbf{#1}}}}
\def\%#1{{\mbox{\boldmath $#1$}}}
\def\=#1{{\overline{\overline{\mathsf #1}}}}

\def\SE{{\mathbb E}}
\def\SF{{\mathbb F}}

\def\*{^{\displaystyle*}}

\def\.{\cdot}
\def\x{\times}

\def\Ra{\Rightarrow}

\def\l#1{\label{eq:#1}}
\def\r#1{(\ref{eq:#1})}
\def\am{\left(\begin{array}{c}}
\def\amm{\left(\begin{array}{cc}}
\def\ammm{\left(\begin{array}{ccc}}
\def\ammmm{\left(\begin{array}{cccc}}
\def\a{\end{array}\right)}

\def\A{\alpha}
\def\B{\beta}
\def\de{\delta}
\def\De{\Delta}

\def\g{\gamma}
\def\G{\Gamma}

\def\la{\lambda}
\def\La{\Lambda}

\def\o{\omega}

\def\t{\tau}

\def\ve{\%\varepsilon}

\def\tr{{\rm tr }}

\def\W{\wedge}
\def\WW{\displaystyle{{}^\wedge}\llap{${}_\wedge$}}

\def\J{\rfloor}
\def\L{\lfloor}
\def\JJ{\rfloor\rfloor}
\def\LL{\lfloor\lfloor}

\begin{document}
\maketitle

\begin{abstract}
It has been known through some examples that parameters of an electromagnetic medium can be so defined that there is no dispersion equation (Fresnel equation) to restrict the choice of the wave vector of a plane wave in such a medium, i.e., that the dispersion equation is satisfied identically for any wave vector. In the present paper, a more systematic study to define classes of media with no dispersion equation is attempted. The analysis makes use of coordinate-free four-dimensional formalism in terms of multivectors, multiforms and dyadics.
\end{abstract}

\section{Introduction}

Time-harmonic plane waves in linear (bi-anisotropic) electromagnetic media are characterized by dispersion equations \cite{Kong} (or Fresnel equations \cite{Hehl}) restricting the choice of the $\#k$ vector of the plane wave for any given frequency $\o$. In the general case the dispersion equation is of the fourth order. For a restricted class of media the fourth-order equation can be reduced to two second-order equations, in which case the field solutions can be decomposed in two independent sets obeying certain polarization properties \cite{Balakin,Obukhov,Deco,Dahl1}. A simple example of such a medium is the one defined by uniaxial permittivity and permeability dyadics. In a still more restricted case, e.g., that of isotropic media, the two equations coincide to a single second-order equation, whence the medium appears free of birefringence \cite{Lammer,Itin05,AFLB,Dahl2}. Finally, one can define media for which the dispersion equation is satisfied identically. In such a case, the medium imposes no restriction on the choice of the $\#k$ vector of the plane wave. It appears that this property has been widely neglected in the literature. However, media belonging to this cathegory have recently emerged in various studies \cite{IB,IBDB,P,Alberto}. Interestingly, it has been shown that novel types of boundary conditions may arise at the interface of some media of this kind \cite{IBDB,ICEAA1,ICEAA2}. It is the purpose of the present paper to define, in a more systematic manner, classes of media in which fields are not restricted by a dispersion equation.

Because of its compactness, the analysis will apply the four-dimensional differential-form formalism of reference \cite{Deschamps} empowered by coordinate-free dyadic representations for linear mappings \cite{Difform}. The source-free Maxwell equations can be expressed in simple form as 
\ea \#d\W\%\Phi &=&0, \l{MaxPhi}\\
    \#d\W\%\Psi &=&0, \l{MaxPsi}\fa
where the electromagnetic two-forms $\%\Phi,\%\Psi\in\SF_2$ are defined by
\ea \%\Phi &=& \#B + \#E\W\ve_4,\\
    \%\Psi &=& \#D - \#H\W\ve_4, \fa
in terms of the spatial 3D one-forms $\#E,\#H\in\SF_1$ and two-forms $\#B,\#D$. The temporal one-form
\e \ve_4 = \#d\t,\ \ \ \ \t=ct, \f
can be completed by spatial one-forms $\ve_1,\ve_2,\ve_3$ to a basis of one-forms.  The complementary basis vectors $\#e_1,\#e_2,\#e_3,\#e_4\in\SE_1$ satisfy $\#e_i|\ve_j=\de_{ij}$. Details of the notation and operational rules applied in this paper can be found in \cite{Difform}. For similar equations in tensorial notation one may visit the references \cite{Hehl,Alberto}.

\section{Dispersion dyadic}

A plane wave can be defined by fields depending on the space-time vector $\#x=\#r+ \#e_4\t$ as
\ea \%\Phi(\#x) &=& \%\Phi f(\%\nu|\#x),\\
    \%\Psi(\#x) &=& \%\Psi f(\%\nu|\#x). \fa
Here, $f(.)$ is any differentiable scalar function and $\%\nu\in\SF_1$ is the wave one-form which contains the wave vector $\#k$ and the frequency $\o$ of the Gibbsian 3D vector notation. For a plane wave the Maxwell equations \r{MaxPhi} and \r{MaxPsi} imply
\ea \%\nu\W\%\Phi&=&0,\l{nuPhi}\\
 \%\nu\W\%\Psi &=& 0. \l{nuPsi}\fa
Applying the identity \cite{Difform}
\e \#a\J(\%\nu\W\%\Phi)= \%\nu\W(\#a\J\%\Phi) + (\#a|\%\nu)\%\Phi, \f
and similarly for $\%\Psi$, valid for any vector $\#a$, assuming $\#a|\%\nu\not=0$, we can express the field two-forms in terms of potential one-forms $\%\phi,\%\psi$ as
\ea \%\Phi&=&\%\nu\W\%\phi,\ \ \ \ \ \%\phi = -\frac{\#a\J\%\Phi}{\#a|\%\nu}, \\
    \%\Psi&=&\%\nu\W\%\psi,\ \ \ \ \ \%\psi = -\frac{\#a\J\%\Psi}{\#a|\%\nu}. \fa

A linear medium can be represented in terms of a linear mapping between the electromagnetic two-forms. Let us assume that the medium is homogeneous and time-invariant, i.e., independent of $\#x$. In terms of a bidyadic $\=M\in\SF_2\SE_2$, mapping two-forms to two-forms, one writes
\e \%\Psi = \=M|\%\Phi. \l{PsiM}\f
Another, equally valid representation, makes use of another bidyadic $\=N\in\SF_2\SE_2$ as
\e \%\Phi = \=N|\%\Psi. \l{PhiN}\f
$\=M$ and $\=N$ may be expanded in any given bivector basis and two-form basis as $6\x6$ matrices, each involving 36 scalar parameters in the general case \cite{Post}. When the two bidyadics are of full rank, they are inverses of each other, while in the converse case only one of the two representations \r{PsiM} and \r{PhiN} may exist. The medium equations can be equally well expressed in terms of modified medium bidyadics $\=M_m,\=N_m\in\SE_2\SE_2$ as bivector equations
\ea \#e_N\L\%\Psi &=& \=M_m|\%\Phi, \l{Mm}\\
\#e_N\L\%\Phi &=& \=N_m|\%\Psi, \l{Nm}\fa
in terms of the quadrivector $\#e_N=\#e_{1234}$ and
\e \=M_m = \#e_N\L\=M,\ \ \ \ \ \=N_m=\#e_N\L\=N \f
mapping two-forms to bivectors.

According to Hehl and Obukhov \cite{Hehl}, the general medium bidyadic can be decomposed in three components as
\e \=M = \=M_1 + \=M_2+ \=M_3, \f
respectively called the principal, skewon and axion parts of $\=M$. The axion part is a multiple of the unit bidyadic, $\=M_3=M_3\=I{}^{(2)T}$, while $\=M_1$ and $\=M_2$ are trace free. The skewon part is characterized by the modified bidyadic $\=M_{m2}=\#e_N\L\=M_2$ which is the antisymmetric part of $\=M_m$, while the principal modified bidyadic $\=M_{m1}=\#e_N\=M_1$ is symmetric. The other medium bidyadic $\=N$ can be given a similar decomposition.

Condition \r{nuPsi} defines an equation for the potential one-form $\%\phi$ as
\e \%\nu\W\%\Psi = (\%\nu\W\=M\L\%\nu)|\%\phi=0. \l{cond}\f
The dyadic in brackets belongs to the space $\SF_3\SE_1$ mapping one-forms to three-forms. An equivalent equation can be formed in terms of the dyadic $\=D(\%\nu)\in\SE_1\SE_1$ mapping one-forms to vectors 
\e \=D(\%\nu) = \#e_N\L(\%\nu\W\=M\L\%\nu)= -\%\nu\J(\#e_N\L\=M)\L\%\nu = \=M_m\LL\%\nu\%\nu, \l{D}\f
and called the dispersion dyadic, as 
\e \=D(\%\nu)|\%\phi=0.\f
From the form of \r{D} it follows that the dispersion dyadic also satisfies 
\e \=D(\%\nu)|\%\nu=0\f
for any $\%\nu$. Assuming non-vanishing field two-forms $\%\Phi=\%\nu\W\%\phi\not=0$, the one-forms $\%\phi$ and $\%\nu$ are linearly independent, whence the rank of the dispersion dyadic is demanded to be less than 3. Thus, $\=D(\%\nu)$ must satisfy the equation \cite{Difform}
\e \=D{}^{(3)}(\%\nu)= \frac{1}{6}\=D(\%\nu)\WW\=D(\%\nu)\WW\=D(\%\nu)= 0, \l{D3nu}\f
for any $\%\nu$ characterizing a plane wave in such a medium. One can show that we can expand
\e \=D{}^{(3)}(\%\nu)=(\#e_N\L\%\nu)(\#e_N\L\%\nu)D(\%\nu)=0, \f
so that \r{D3nu} is actually equivalent to a scalar equation which can be given the following explicit form \cite{Disp}, 
\ea D(\%\nu) &=& \frac{1}{3}\ve_N\ve_N||(\=M_m\WW\=D{}^{(2)}(\%\nu)) \nonumber\\  &=&\frac{1}{6}\ve_N\ve_N||(\=M_m\WW(\%\nu\%\nu\JJ(\=M_m\WW(\%\nu\%\nu\JJ\=M_m))))=0. \l{Disp}\fa
\r{Disp} (or its equivalent dyadic form \r{D3nu}) is called the dispersion equation, because for time-harmonic fields it defines the relation between the wave number and the frequency of the wave. Alternatively, it is called the Fresnel equation. Obviously, \r{Disp} is of the fourth order in $\%\nu$. Other forms for the dispersion equation are found in \cite{Obu,Rubilar,Hehl,Itin,Alberto} 

Starting from the medium condition \r{PhiN}, a similar equation in terms of the bidyadic $\=N_m$ is obtained,
\e D'(\%\nu) = \frac{1}{6}\ve_N\ve_N||(\=N_m\WW(\%\nu\%\nu\JJ(\=N_m\WW(\%\nu\%\nu\JJ\=N_m))))=0. \l{DispN}\f
When the medium bidyadics $\=M,\=N$ are of full rank, the ensuing two dispersion equations must have the same solutions since they correspond to the same plane wave \cite{Dahl2}. In particular, if one of them is satisfied identically for all $\%\nu$, so must be the other one. When the rank of $\=M$ is less than six, the bidyadic $\=N$ and the equation \r{DispN} do not exist, and conversely.

\section{Media with no dispersion equation}

To define media with no dispersion equation, either \r{D3nu} or \r{Disp} must be identically valid for any one-form $\%\nu$. Two examples have emerged in the past, the skewon-axion media (IB-media) \cite{Hehl,IB} and the P-media \cite{P,Alberto}. Actually these two media will also come out through the present analysis, added by a third one denoted by case 1 below.

Starting from \r{D3nu}, requiring that the dispersion dyadic be of rank less than three, it follows that there must exist four vectors in terms of which we can write
\e \=D(\%\nu) = \#a\#c + \#b\#d. \l{Dacbd}\f
This corresponds to the case of planar dyadics in the 3D Gibbsian formalism \cite{Gibbs,Methods}.

Since the left side of \r{Dacbd} is a quadratic function of $\%\nu$, so must be the right side. This gives us a few possibilities.
\begin{enumerate}
\item Each of the four vectors is a linear function of $\%\nu$
\item $\#a$ and $\#b$ are quadratic functions while $\#c$ and $\#d$ do not depend on $\%\nu$
\item $\#a$ and $\#d$ are quadratic functions while $\#b$ and $\#c$ do not depend on $\%\nu$
\item $\#a$ is a quadratic function while $\#b$ and $\#d$ are linear functions and $\#c$ does not depend on $\%\nu$
\end{enumerate}
Other possibilities do not seem to bring any new solutions because the dispersion equation \r{Disp} is invariant to replacing the modified medium bidyadic $\=M_m$ by its transpose $\=M{}_m^T$. In fact, this implies replacing the dispersion dyadic $\=D(\%\nu)$ by its transpose $\=D{}^T(\%\nu)$ and $\=D{}^{(3)}(\%\nu)$ by $\=D{}^{(3)T}(\%\nu)$. Let us consider these four cases separately.

\subsection{Case1: $\#a,\#b,\#c$ and $\#d$ are linear functions of $\%\nu$}

Because the dispersion dyadic satisfies $\=D(\%\nu)|\%\nu =\%\nu|\=D(\%\nu) =0$ for all $\%\nu$, the four vectors must satisfy
\e \#a(\#c|\%\nu) + \#b(\#d|\%\nu)=0,\ \ \ \ (\%\nu|\#a)\#c + (\%\nu|\#b)\#d=0. \f
Assuming 
\e \=D{}^{(2)}= (\#a\W\#b)(\#c\W\#d)\not=0, \l{D2not0} \f 
which requires that the dispersion dyadic be of rank 2, the vectors $\#a,\#b$ must be linearly independent and so must $\#c,\#d$, whence we obtain
\e \#a|\%\nu= \#b|\%\nu=\#c|\%\nu=\#d|\%\nu=0, \f
for all $\%\nu$. This requires that there must exist bivectors $\#A,\#B,\#C,\#D$ such that we can express 
\e \#a=\#A\L\%\nu,\ \ \ \ \#b=\#B\L\%\nu,\ \ \ \ \#c=\#C\L\%\nu,\ \ \ \ \#d=\#D\L\%\nu, \l{ABCD}\f
and
\e \=D(\%\nu) = \=M_m\LL\%\nu\%\nu= (\#A\#C+\#B\#D)\LL\%\nu\%\nu. \f
Because  
\e (\=M_m -(\#A\#C+\#B\#D))\LL\%\nu\%\nu=0 \f
must be valid for all $\%\nu$, applying the result given in Appendix 1, the modified medium bidyadic $\=M_m$ must be of the form
\e \=M_m =\#A\#C+\#B\#D + \A\#e_N\L\=I{}^{(2)T}, \l{case1}\f
for some scalar $\A$. Equivalently, the medium bidyadic $\=M$ must be of the form
\e \=M =\%\Pi\#C+\%\La\#D + \A\=I{}^{(2)T},  \l{case1a}\f
with $\%\Pi=\ve_N\L\#A$ and $\%\La=\ve_N\L\#B$. The expression \r{case1a} defines one possible class of media in terms of two bivectors, two two-forms and a scalar, for which the dispersion equation is satisfied identically. The medium bidyadic $\=N=\=M{}^{-1}$ has a similar form, which can be shown by forming the inverse bidyadic of \r{case1a}, details of which are given in Appendix 2. 

In the special case $\=D{}^{(2)}(\%\nu)=0$ for all $\%\nu$, \r{case1} is reduced to the form
\e \=M_m =\#A\#C+ \A\#e_N\L\=I{}^{(2)T}, \l{case1'}\f
while requiring $\=D(\%\nu)=0$ leads to $\=M=M\=I{}^{(2)T}$, i.e., a medium with just an axion component. One may note that the medium defined by \r{case1} includes any Q-medium \cite{Difform}, defined by 
\e \=M_m = M\=Q{}^{(2)}, \f
when the dyadic $\=Q\in\SE_1\SE_1$ is antisymmetric. In fact, because such a dyadic can be expressed as $\=Q=\#A\L\=I{}^T$, where $\#A$ is some bivector, expanding 
\ea \=M_m &=&M(\#A\L\=I{}^T)^{(2)} \nonumber\\
&=& M(\#A\#A - \frac{1}{2}\ve_N|(\#A\W\#A)\=I{}^{(2)T}), \fa  
yields a special case of \r{case1'}.

One may note that the condition \r{D2not0} requires that none of the four bivectors in \r{case1} is simple. In fact if, for example, we assume that $\#A$ simple, whence we can express it in terms of two vectors as $\#A=\#a_1\W\#a_2$, we have $\#a=\#A\L\%\nu=(\#a_2|\%\nu)\#a_1-(\#a_1|\%\nu)\#a_2=0$ for any $\%\nu$ satisfying $\#a_1|\%\nu=0$ and $\#a_2|\%\nu=0$. Thus, \r{D2not0} is not valid for one-forms $\%\nu$ taken from the corresponding 2D subspace. In such a case the dispersion dyadic $\=D(\%\nu)$ is of rank 1 for a 2D subspace of $\%\nu$ and of rank 2 for other one-forms $\%\nu$. Since media defined by \r{case1} have no dispersion equation regardless of one or more of the bivectors being simple, dispersion dyadics $\=D(\%\nu)$ of variable rank are included in addition to those of rank 2 or rank 1 for all $\%\nu$. One may conjecture that \r{case1} represents all solutions for case 1 media.

\subsection{Case 2: $\#a$ and $\#b$ are quadratic, while $\#c$ and $\#d$ are independent of $\%\nu$}

Let us now assume that in \r{Dacbd} the vectors $\#c$ and $\#d$ are independent of $\%\nu$, whence $\#a$ and $\#b$ must be quadratic functions of $\%\nu$. In this case it appears preferable to start from the representation
\e \%\nu\W\=M\L\%\nu = -\ve_N\L\=D(\%\nu) = \%\G\#c +\%\De\#d, \l{veNab} \f 
where the two three-forms 
\e \%\G=-\ve_N\L\#a,\ \ \ \%\De=- \ve_N\L\#b,\ \ \ \in\SF_3 \l{GD}\f
are now quadratic functions of $\%\nu$. Since any two linearly independent three-forms can be interpreted as belonging to a basis of three-forms, which can be constructed in terms of a basis of one-forms, say $\%\A,\%\B,\%\g,\%\de$, we can express 
\e \%\G = \%\A\W\%\B\W\%\g,\ \ \ \ \%\De=\%\A\W\%\B\W\%\de. \f
Thus, \r{veNab} can be written as
\e \%\nu\W\=M\L\%\nu = \%\A\W\%\B\W(\%\g\#c +\%\de\#d), \l{veNAB}\f
or, more generally, as
\e \%\nu\W\=M\L\%\nu= \%\A\W\%\B\W\=P{}^T. \l{nuABB} \f
Since terms of the form $\%\A\#a'+ \%\B\#b'$ can be added to the dyadic in brackets in \r{veNAB}, we may well assume that the dyadic $\=P\in\SE_1\SF_1$ in \r{nuABB} is of full rank. The special case corresponding to $\=D{}^{(2)}(\%\nu)=0$, i.e., when the bracketed dyadic is reduced by assuming $\%\de\#d=0$, will be considered separately. 

Before proceeding, let us briefly check that when $\=M$ satisfies a relation of the form \r{nuABB}, the dispersion equation is satisfied identically. The dispersion dyadic can be expressed in the form
\e \=D(\%\nu) = \#e_N\L(\%\nu\W\=M\L\%\nu) = \#D\L\=P{}^T = (\#D\L\=I{}^T)|\=P{}^T, \l{DABPT}\f
where the bivector $\#D\in\SE_2$ is defined by
\e \#D = \#e_N\L(\%\A\W\%\B), \f
and $\#D\L\=I{}^T\in\SE_1\SE_1$ is an antisymmetric dyadic. The bivector $\#D$
is simple because it satisfies
\e \ve_N|(\#D\W\#D) = \ve_N|((\#e_N\L(\%\A\W\%\B))\W(\#e_N\L(\%\A\W\%\B))) = \#e_N|(\%\A\W\%\B\W\%\A\W\%\B)=0. \f
Since a simple bivector can be expressed in terms of two vectors as $\#D=\#d_1\W\#d_2$, the antisymmetric dyadic satisfies
\ea (\#D\L\=I{}^T)^{(3)} &=& ((\#d_1\W\#d_2)\L\=I{}^T)^{(3)}=(\#d_2\#d_1-\#d_1\#d_2)^{(3)}\nonumber\\
&=& \frac{1}{3}(\#d_2\#d_1-\#d_1\#d_2)\WW(\#d_1\#d_2\WW\#d_2\#d_1)=0. \l{D3baab}\fa
From this and the rule $(\=A|\=B)^{(p)} = \=A{}^{(p)}|\=B{}^{(p)}$ it follows that the dispersion equation 
\e \=D{}^{(3)}(\%\nu) = (\#D\L\=I{}^T)^{(3)}|\=P{}^{(3)T}=0 \l{D3PT} \f
is, indeed, satisfied for any $\%\nu$ and any medium bidyadic $\=M$ satisfying \r{nuABB}. 

Returning to \r{nuABB}, two conditions for the quantities on its right-hand side are obtained by noting that the left-hand side vanishes when multiplying by $|\%\nu$ from the right or by $\%\nu\W$ from the left. These lead to the respective conditions 
\e \%\A\W\%\B\W(\=P{}^T|\%\nu)=0, \l{ABP}\f
and
\e \%\nu\W\%\A\W\%\B\W\=P{}^T= 0. \l{nuWABB}\f
Condition \r{ABP} requires that the three one-forms $\%\A, \%\B$ and $\=P{}^T|\%\nu$ be linearly dependent, whence they must satisfy a relation of the form
\e A\%\A+ B\%\B + C(\=P{}^T|\%\nu)=0. \f 
For $C=0$ we have $\%\A\W\%\B=0$ and $\=D(\%\nu)=0$ corresponding to the axion medium. For $C\not=0$ we can choose $C=-1$, whence the relation can be rewritten as 
\e \=P{}^T|\%\nu=A\%\A + B\%\B. \f
Omitting vanishing of both $A$ and $B$ which would again yield the axion medium, without sacrifying generality ($\%\A$ and $\%\B$ appear interchangeable at this stage), we may set $B=1$, whence
\e \%\B = \=P{}^T|\%\nu -A\%\A. \f
Inserting this in \r{nuABB}, the condition becomes
\e \%\nu\W\=M\L\%\nu = \%\A\W(\=P{}^T|\%\nu)\W\=P{}^T. \l{BAPT}\f
Since the right-hand side of \r{BAPT} must be a quadratic function of $\%\nu$, it appears that $\=P$ must be independent of $\%\nu$. Thus, $\%\A$ must be a linear function of $\%\nu$ and it can be expressed as
\e \%\A = \=B{}^T|\%\nu, \f
in terms of some dyadic $\=B\in\SE_1\SF_1$. Inserting this in \r{BAPT} yields the representation
\e \%\nu\W\=M\L\%\nu = (\=B{}^T|\%\nu)\W(\=P{}^T|\%\nu)\W\=P{}^T, \l{ABPT1}\f

Applying now the second condition \r{nuWABB}, i.e., multiplying \r{ABPT1} by $\%\nu\W$ from the left, leads to the condition
\e \%\nu\W(\=B{}^T|\%\nu)\W(\=P{}^T|\%\nu)\W\=P{}^T=0, \l{cond1P}\f
valid for any one-form $\%\nu$. Recalling that $\=P$ was assumed to be a dyadic of full rank, we can multiply \r{cond1P} by $|\=P{}^{-1T}$ from the right. The condition is then reduced to
\e \%\nu\W(\=B{}^T|\%\nu)\W(\=P{}^T|\%\nu)\W\=I{}^T=0\ \ \ \Ra\ \ \ \%\nu\W(\=B{}^T|\%\nu)\W(\=P{}^T|\%\nu)=0. \l{cond1}\f
Since the three one-forms are linearly dependent for all $\%\nu$, the dyadics must be related as 
\e A\=I + B\=B + C\=P=0, \f
for some scalars $A,B,C$. This leaves us the following three possibilities:
\begin{enumerate}
\item $A=0$ which implies $B\=B{}^T|\%\nu+ C\=P{}^T|\%\nu=0$. Since $\%\nu\W\=M\L\%\nu=0$, we must have 
\e \=M=\A\=I{}^{(2)T}. \f
\item $A\not=0$ and $B=0$ implies $\=P = -(A/C)\=I$ and $\%\nu\W\=M\L\%\nu = (A/C)\%\nu\W(\=B\WW\=I)^T\L\%\nu$. In this case we must have
\e  \=M = (\=B\WW\=I)^T + \A\=I{}^{(2)T}, \l{skwax}\f
\item $A\not=0$ and $B\not=0$ implies $\=B = -(A/B)\=I - (C/B)\=P$. In this case we must have
\e \%\nu\W\=M\L\%\nu = -(A/B)\%\nu\W\=P{}^T|\%\nu\W\=P{}^T = -(A/B)\%\nu\W\=P{}^{(2)T}\L\%\nu, \f
and
\e \=M =  \=P{}^{(2)T} + \A\=I{}^{(2)T}. \l{Pax} \f
\end{enumerate}
In \r{skwax} and \r{Pax} the coefficients $A,B,C$  have been suppressed. Thus, in case 2 the media with no dispersion equation fall into two classes, those of skewon-axion media \cite{IB} and P-axion media while axion media are special cases of both of the two. The class of P-axion media is a generalization of that of P-media \cite{P}, with an added axion term. 

Actually, all of the above medium bidyadics can be expressed under the single, more general, form as 
\e \=M = A\=B{}^{(2)T} + B(\=B\WW\=I)^T + C\=I{}^{(2)T} \l{case2} \f
where $A,B,C$ are any scalar coefficients.  In fact, \r{case2} yields the skewon-axion medium for $A=0$ and the P-axion medium for $A\not=0$. For $A\not=0$ this can be seen by defining new symbols as
\e \=P = \=B + \frac{B}{A}\=I,\ \ \ M=A,\ \ \ \A=C-\frac{B^2}{A}. \f
Let us finally test whether \r{case2} really satisfies the dispersion equation \r{D3nu} for any one-form $\%\nu$ by expanding the dispersion dyadic as
\ea \=D(\%\nu) &=& -\#e_N\L(\%\nu\W\=M\L\%\nu) \nonumber\\ 
&=& -\#e_N\L(A\%\nu\W\=B{}^{(2)T}\L\%\nu + B\%\nu\W(\=B\WW\=I)^T\L\%\nu) \nonumber\\
&=& -\#e_N\L(A\%\nu\W(\=B{}^T|\%\nu)\W\=B{}^T + B\%\nu\W(\=B{}^T|\%\nu)\W\=I{}^T) \nonumber\\
&=& (-\#e_N\L(\%\nu\W(\=B{}^T|\%\nu)))\L(A\=B+B\=I)^T  \nonumber\\
&=& (\#A\L\=I{}^T)|(A\=B+B\=I)^T.  \fa 
Here we have denoted $\#A=-\#e_N\L(\%\nu\W(\=B{}^T|\%\nu))$, which is a simple bivector satisfying $\#A\W\#A=0$. Now the dispersion equation \r{D3nu} becomes
\e \=D{}^{(3)}(\%\nu)=  (\#A\L\=I{}^T)^{(3)}|(A\=B+B\=I)^{(3)T}=0, \f
which is valid for the same reason as \r{D3baab}. It is remarkable that this is valid for any dyadic $\=B$ which means that the assumption of full-rank dyadic $\=P$ can be relaxed. 

The previous analysis assumed $\=D{}^{(2)}(\%\nu)\not=0$ for all $\%\nu$. The converse case can be handled by replacing \r{Dacbd} by
\e \=D(\%\nu) = \#a\#c, \f
where $\#a$ is a quadratic function of $\%\nu$ and $\#c$ is independent of $\%\nu$. From  
\e \=D(\%\nu)|\%\nu= \#a(\#c|\%\nu)=0, \f
satisfied for all $\%\nu$, we must have either $\#a=0$ or $\#c=0$, both of which lead to $\=D(\%\nu)=0$. This, again, corresponds to the axion medium case which is included the solution \r{case2}.

\subsection{Case 3: $\#a$ and $\#d$ are quadratic in $\%\nu$ while $\#b$ and $\#c$ are independent of $\%\nu$ }

In this case we assume that in the dispersion dyadic representation \r{Dacbd} the vectors $\#c$ and $\#b$ are independent of $\%\nu$. From  
\e \#b\W\=D(\%\nu)|\%\nu = \#b\W(\#a\#c+ \#b\#d)|\%\nu = (\#b\W\#a)(\#c|\%\nu)=0, \f
valid for all $\%\nu$, we conclude that either $\#c=0$ or $\#b\W\#a=0$, whence $\#a$ must be of the form $\#a=(\=A'||\%\nu\%\nu)\#b$ for some dyadic $\=A'\in\SE_1\SE_1$.  Similarly, we must have either $\#b=0$ or $\#d=(\=D'||\%\nu\%\nu)\#c$. Thus, the dispersion dyadic of case 3 coincides with that of case 2, or its transpose, with $\=D{}^{(2)}(\%\nu)=0$ for all $\%\nu$.

\subsection{Case 4: $\#a$ is quadratic, $\#b$ and $\#d$ are linear and $\#c$ is independent of $\%\nu$}

In this final case from
\e \=D(\%\nu)|\%\nu = \#a(\#c|\%\nu) + \#b(\#d|\%\nu)=0, \f
assuming $\#a\W\#b\not=0$ we obtain $\#c|\%\nu=0$ and $\#d|\%\nu=0$ for all $\%\nu$. This requires $\#c=0$ and $\#d=\#D\L\%\nu$, whence the dispersion dyadic must be of the form
\e \=D(\%\nu) = \#B\#D\LL\%\nu\%\nu \l{case4D}\f
for some bivectors $\#B,\#D$ This leads to case 1 solution \r{case1}. Alternatively, from $\#a\W\#b=0$ we obtain the relation $\#a = \la\#b = (\#a'|\%\nu)\#b$, since $\la$ must be a linear function of $\%\nu$. In this case the dispersion dyadic becomes
\e \=D(\%\nu) = \#b((\#a'|\%\nu)\#c+ \#d), \f
which reduces to the form \r{case4D}. In conclusion, case 4 coincides with the special case 1 satisfying $\=D{}^{(2)}(\%\nu)=0$ for all $\%\nu$. Thus, the corresponding modified medium bidyadic has the form of \r{case1'}.

\section{Discussion}

The previous analysis shows us that the possible medium bidyadic $\=M$, in which a plane wave is not rectricted by a dispersion equation, is either of the general form \r{case1} or \r{case2}.  Starting from the representation \r{PhiN}, similar expressions can be obtained for the bidyadic $\=N$, which equals $\=M{}^{-1}$ when the two bidyadics are of full rank. It is shown in Appendix 2 that both of these representations lead to the same class of media for case 1. Let us now study whether the same property is valid for case 2. It is known from previous work that for pure axion, skewon and P-medium bidyadics the inverse bidyadics, when they exist, have the same axion, skewon or P-medium character \cite{IB,P,Alberto}. So, one may wonder whether this property will be valid for skewon-axion and P-axion media as well. Since there are no analytic expressions known to these authors for the inverse of general skewon-axion or P-axion bidyadics, the problem must be considered through some known properties.

It is known that the modified medium bidyadic of a P-medium 
\e \=M_m = M\#e_N\L\=P{}^{(2)T}, \l{MmP}\f
for some dyadic $\=P\in\SE_1\SF_1$ satisfies the quadratic equation \cite{Schuller,Deco,Dahl3}
\e \=M{}^T_m\.\=M_m = P\#e_N\L\=I{}^{(2)T},\ \ \ \  P=M^2\tr\=P{}^{(4)}. \l{MmM}\f
The natural dot product between two bidyadics $\#A,\#B$ or two two-forms $\%\Phi,\%\Psi$ is defined by
\ea \#A\.\#B &=& \#A|(\ve_N\L\=I{}^{(2)}|\#B = \#B\.\#A,\\
    \%\Phi\.\%\Psi &=& \%\Phi|(\#e_N\L\=I{}^{(2)T}|\%\Psi = \%\Psi\.\%\Phi. \fa
The converse is not necessarily true, since, as was shown in \cite{Deco}, a solution of \r{MmM} can alternatively be a bidyadic of a Q-medium, which has the form \cite{Difform}
\e \=M_m = M\=Q{}^{(2)}, \f
for some dyadic $\=Q\in\SE_1\SE_1$. However, as is shown in Appendix 3, if a solution of \r{MmM} is of full rank $(P\not=0)$ and belongs to case 2, it must be a P-medium bidyadic, since Q-medium bidyadics possessing no dispersion equation fall necessarily into case 1. 

From \r{MmM} it follows that the bidyadic of a P-axion medium 
\e \=M_m = M\#e_N\L\=P{}^{(2)T} + \A\#e_N\L\=I{}^{(2)T}, \f
satisfies the bidyadic second-order equation
\e \=M{}^T_m\.\=M_m -\A(\=M{}^T_m+\=M_m) = (P-\A^2)\#e_N\L\=I{}^{(2)T}. \l{MTM}\f
Let us multiply \r{MTM} by $\=M{}_m^{-1T}|$ from the left and by $|\=M{}_m^{-1}$ from the right, whence we arrive at
\e \ve_N\L\=I{}^{(2)} -\A(\=M{}_m^{-1T}+\=M{}_m^{-1})=(P-\A^2)\=M{}_m^{-1T}|(\#e_N\L\=I{}^{(2)T})|\=M{}_m^{-1} . \f
For $P\not=\A^2$ we can find the equation corresponding \r{MTM} for the bidyadic $\=N_m=\#e_N\#e_N\LL\=M{}_m^{-1}$, 
\e \=N{}^T_m\.\=N_m+ \frac{\A}{P-\A^2}(\=N{}^T_m+\=N_m)=\frac{1}{P-\A^2}\#e_N\L\=I{}^{(2)T}. \l{NTN}\f
Since \r{NTN} is of the same form as \r{MTM}, we may conclude that, in the general case, the inverse of a P-axion medium is another P-axion medium.

In the special case when P-axion medium bidyadic \r{MmM} is restricted by $P=\A^2$, i.e., for 
\e \=M_m = M\#e_N\L\=P{}^{(2)T} + \A \#e_N\L\=I{}^{(2)T},\ \ \ \ \A=\pm\sqrt{P}, \l{MmMP}\f
with either sign, \r{NTN} reduces to
\e \A(\=N{}^T_m+\=N_m)=\#e_N\L\=I{}^{(2)T},\f
which is satisfied by bidyadics of the form
\e \=N_m = \=A + \frac{1}{2\A}\#e_N\L\=I{}^{(2)T}, \f
where $\=A$ is any antisymmetric bidyadic. From this we conclude that the inverse of a special P-axion medium \r{MmMP} is a skewon-axion medium. This works also backwards so that the inverse of a skewon-axion medium is of the special P-axion medium form \r{MmMP}. This is, however, not valid for $\A=0$, which corresponds to the pure skewon medium whose inverse is another pure skewon medium \cite{Skewon}, while for $P=0$ the bidyadic $\=M_m$ in \r{MmMP} does not have any inverse. The result is shown in Table \ref{tab:table}.

\begin{table}[ht]
	\centering
		\begin{tabular}{|c|c|}
			\hline
			${\overline{\overline{\mathsf M}}}$ & ${\overline{\overline {\mathsf N}}}$\\
			\hline
			axion & axion\\
			skewon & skewon \\
			P-medium & P-medium\\
			skewon-axion & special P-axion\\
					special P-axion & skewon-axion\\
				general P-axion & general P-axion\\
				\hline
		\end{tabular}
	\caption{Relations between various case 2 medium bidyadics. }
	\label{tab:table}
\end{table}

It may be of interest to note that both case 1 and case 2 classes of media with no dispersion equation are closed in affine transformations. Let us consider a full-rank dyadic $\=A\in\SE_1\SF_1$ mapping vectors as $\#x_a=\=A|\#x$, bivectors as $\#A_a = \=A{}^{(2)}|\#A$ and medium bidyadics as \cite{Difform}
\e \=M_a = \=A{}^{(-2)T}|\=M|\=A{}^{(2)T}. \f
Applying this transformation, case 1 medium bidyadics are mapped as
\ea \=M_a &=&  \=A{}^{(-2)T}|(\%\Pi\#C+\%\De\#D+\A\=I{}^{(2)T})|\=A{}^{(2)T}\nonumber\\
&=&  \%\Pi_a\#C_a+\%\De_a\#D_a+\A\=I{}^{(2)}, \fa
with
\e \%\Pi_a = \=A{}^{(-2)T}|\%\Pi,\ \ \ \ \%\De_a = \=A{}^{(-2)T}|\%\De, \f
\e \#C_a = \=A{}^{(2)}|\#C,\ \ \ \ \#D_a = \=A{}^{(2)}|\#D, \f
and case 2 medium bidyadics are mapped as
\ea \=M_a &=&  \=A{}^{(-2)T}|(A\=B{}^{(2)T}+ B(\=B\WW\=I)^T + C\=I{}^{(2)T}|\=A{}^{(2)T}\nonumber\\
&=& A\=B{}_a^{(2)T}+ B(\=B_a\WW\=I)^T + C\=I{}^{(2)T}, \fa
with
\e \=B_a = \=A|\=B|\=A{}^{-1}. \f
Thus, in both cases, the form of the medium bidyadic remains the same after the transformation.

\section{Conclusion}

This paper concerns the problem of defining classes of electromagnetic media in which plane waves are not restricted by a dispersion equation (Fresnel equation). Some media with such a property have recently emerged in various studies. In the analysis four-dimensional formalism in terms of multivectors, multiforms and dyadics was applied as based on reference \cite{Difform}.

Requiring that the dispersion dyadic $\=D(\%\nu)$ satisfy the condition $\=D{}^{(3)}(\%\nu)=0$ for all possible wave one-forms $\%\nu$ the problem was split in four cases which were analyzed separately. Case 1 and case 2 were shown to yield medium bidyadics $\=M$, respectively defined by \r{case1a} and \r{case2}, with the desired property, while cases 3 and 4 could be reduced to cases 1 and 2 with no new solutions. Previously known media showing no dispersion equation were seen to be special cases of the results obtained here. In particular, the Q-medium with antisymmetric $\=Q$ dyadic belongs to the class of case 1 media \r{case1a} while the skewon-axion medium and P-medium are both special cases of the class of case 2 media. When considering the problem in terms of the alternative medium bidyadic $\=N$, no new solutions could be found in any of the cases considered.

\section*{Appendix 1}

Consider the following problem: find the most general bidyadic $\=M\in\SF_2\SE_2$ satisfying
\e \%\A\W\=M\L\%\A=0 \l{AMA}\f
for all possible one-forms $\%\A$. The same problem is solved in \cite{Dahl2} (Proposition 3.1). 

Applying the identity \cite{Difform}
\e \#a\J(\%\A\W\%\Pi) = \%\A\W(\#a\J\%\Pi) + (\#a|\%\A)\%\Pi \f
valid for any one-form $\%\A$, two-form $\%\Pi$ and vector $\#a$, by choosing $\#a$ so that $\#a|\%\A\not=0$, we can write from
\e \#a\J(\%\A\W\=M\L\%\A) = \%\A\W(\#a\J\=M\L\%\A) + (\#a|\%\A)\=M\L\%\A=0, \f
the representation
\e \=M\L\%\A = \%\A\W\=A{}^T,\ \ \ \ \=A{}^T= -\frac{\#a\J\=M\L\%\A}{\#a|\%\A}, \f
in terms of some dyadic $\=A\in\SE_1\SF_1$. Since such a dyadic satisfies 
\e \%\A\W\=A{}^T|\%\A = \=M|(\%\A\W\%\A)=0, \f 
we must have
\e \=A{}^T|\%\A = \%\A|\=A= A\%\A. \f
This means that $\%\A$ is a left eigen-one-form of the dyadic $\=A$. Because this must be valid for any one-form $\%\A$, the dyadic is actually a multiple of the unit dyadic,
\e \=A=A\=I,\f
and
\e \=M\L\%\A= A\%\A\W\=I{}^T = A\=I{}^{(2)T}\L\%\A. \f
Since $\=M$ must satisfy
\e (\=M- A\=I{}^{(2)T})\L\%\A=0 \f
for any one-form $\%\A$, by choosing basis one-forms $\%\A=\ve_i$, we finally obtain
$$  \=M- A\=I{}^{(2)T} = (\=M- A\=I{}^{(2)T})|\sum \ve_{ij}\#e_{ij} =\sum( (\=M- A\=I{}^{(2)T})\L\ve_i)|\ve_j\#e_{ij}=0, $$
whence $\=M$ must be a multiple of the unit bidyadic $\=I{}^{(2)T}$.

Multiplying \r{AMA} by $\#e_N\L$ and applying 
\e \#e_N\L(\%\nu\W\=M\L\%\nu) = -\%\nu\J(\#e_N\L\=M)\L\%\nu = \=M_m\LL\%\nu\%\nu, \f
we obtain the following rule: if 
\e \=M_m\LL\%\nu\%\nu=0 \f
is satisfied for all $\%\nu$, the metric bidyadic must be of the form
\e \=M_m = M\#e_N\L\=I{}^{(2)T}. \f

\section*{Appendix 2}

Let us consider a bidyadic 
\e \=M = \%\Pi\#C + \%\La\#D+ \A\=I{}^{(2)}, \l{PiC}\f
defined by two two-forms $\%\Pi,\%\La$, two bivectors $\#C,\#D$ and a scalar $\A$. Let us find the inverse bidyadic by starting from the ansatz
\e \=M{}^{-1} = \%\Pi\#C' + \%\La\#D'+ \A'\=I{}^{(2)}, \l{PiC'}\f
where $\#C,\#D$ and $\A$ are replaced by the unknown quantities $\#C',\#D',\A'$.

Inserting \r{PiC} and \r{PiC'} in 
\e \=M|\=M{}^{-1}=\=I{}^{(2)}, \l{MMinv} \f
and assuming that $\%\Pi$ and $\%\La$ are linearly independent and 
\e \A'=1/\A, \f
we obtain a relation between the unknown and known bivectors as
\e \amm \%\Pi|\#C + \A & \%\La|\#C\\ \%\Pi|\#D & \%\La|\#D+\A\a \am \#C'\\ \#D'\a = -\A'\am \#C\\ \#D\a. \f
This can be solved as
\e \am \#C'\\ \#D'\a = \frac{-1}{\A D}\amm \%\La|\#D + \A & -\%\La|\#C\\ -\%\Pi|\#D & \%\Pi|\#C+\A\a \am \#C\\ \#D\a, \l{C'D'}\f
\e D = (\%\Pi|\#C + \A)(\%\La|\#D+\A)-(\%\La|\#C)(\%\Pi|\#D). \f
It is easy to verify that \r{MMinv} is satisfied by \r{PiC'} with \r{C'D'} substituted. In the case $D=0$ there is no inverse.

For the special case $\#D=0$ in \r{PiC} the result \r{C'D'} is reduced to
\e \#C' = \frac{-1}{\A(\%\Pi|\#C+\A)}\#C. \f

\section*{Appendix 3}

Since \r{MmM} contains P-medium and Q-medium solutions, one may ask how to distinguish whether a given solution $\=M_m$ corresponds to a P-medium or a Q-medium?

Let us assume that the modified medium bidyadic $\=M_m$ satisfies \r{MmM} and study first the Q-solution by making contraction by a one-form $\%\A$ as
\e \=M_m\LL\%\A\%\A = M\=Q{}^{(2)}\LL\%\A\%\A = M(\=Q||\%\A\%\A)\=Q -(\=Q|\%\A)(\%\A|\=Q). \f
Its double-cross square yields
\e (\=M_m\LL\%\A\%\A)^{(2)} = M^2(\=Q||\%\A\%\A)((\=Q||\%\A\%\A)\=Q{}^{(2)}  -\=Q\WW((\=Q|\%\A)(\%\A|\=Q))), \f 
while the double-cross cube becomes
\e (\=M_m\LL\%\A\%\A)^{(3)} = M^3(\=Q||\%\A\%\A)^2((\=Q||\%\A\%\A)\=Q{}^{(3)}  -\=Q{}^{(2)}\WW((\=Q|\%\A)(\%\A|\=Q))). \f 
Applying dyadic rules \cite{Difform}, this can be expressed in the form
\ea (\=M_m\LL\%\A\%\A)^{(3)} &=& M^3(\=Q||\%\A\%\A)^2((\=Q||\%\A\%\A)\=Q{}^{(3)}  -\=Q{}^{(2)}\WW((\=Q|\%\A)(\%\A|\=Q)))\nonumber\\
&=& M^3(\=Q||\%\A\%\A)^2(\ve_N\ve_N||\=Q{}^{(4)})(\#e_N\#e_N\LL\%\A\%\A). \l{M3Q} \fa 

Let us now do the same operations for the P-medium bidyadic.
\ea \=M_m\LL\%\A\%\A &=& M(\#e_N\L\=P{}^{(2)T})\LL\%\A\%\A \nonumber\\
&=& M \#e_N\L(\%\A\W(\=P{}^T|\%\A)\W\=P{}^T). \fa
Denoting the simple bivector by
\e \#e_N\L(\%\A\W(\=P{}^T|\%\A))=\#a\W\#b, \f
we have
\e \=M_m\LL\%\A\%\A = (\#a\W\#b)\L\=P{}^T= \#b(\#a|\=P{}^T) - \#a(\#b|\=P{}^T) \f
Its double-cross square yields
\e (\=M_m\LL\%\A\%\A)^{(2)} = (\#b\W\#a)((\#a|\=P{}^T)\W(\#b|\=P{}^T)),\f
and the double-cross cube vanishes
\e (\=M_m\LL\%\A\%\A)^{(3)} = 0, \l{MmAA}\f
for any one-forms $\%\A$.

Now this could serve as a test for a modified medium bidyadic $\=M_m$ which satisfies \r{MmM} and is of full rank, whence both $\=Q$ and $\=P$ must be of rank 4. If $\=M_m$ satisfies $(\=M_m\LL\%\A\%\A)^{(3)} \not= 0$ for any one-form $\%\A$, it corresponds to a Q-solution. In the converse case, it either corresponds to a P-solution or a special Q-solution with antisymmetric dyadic $\=Q$. $\ve_N\ve_N||\=Q{}^{(4)}=0$ is ruled out by $\=Q$ being of full rank. $\=Q$ antisymmetric implies also $(\=M_m\LL\%\A\%\A)^{(2)} =0$ for all $\%\A$, but the same requirement for the P-medium bidyadic would lead to either $\#a\W\#b=0$ or to $(\#a|\=P{}^T)\W(\#b|\=P{}^T)=(\#a\W\#b)|\=P{}^{(2)T}=0$, which are the same condition for full-rank $\=P$. This condition equals $\%\A\W(\=P{}^T|\%\A)=0$ for all $\%\A$ which is only valid for $\=P$ being a multiple of $\=I$. This last case, again, corresponds to $\=M_m\LL\%\A\%\A=0$ for all $\%\A$, which applied to the Q-medium bidyadic is contrary to the full-rank assumption.

To conclude, we have the following test for the medium bidyadics of Q- and P-media satisfying \r{MmM} when the modified medium bidyadic $\=M_m$ is of full rank.
\begin{itemize}
\item For $(\=M_m\LL\%\A\%\A)^{(3)} \not= 0$ we have a Q-solution
\item For $(\=M_m\LL\%\A\%\A)^{(3)} = 0$ and $(\=M_m\LL\%\A\%\A)^{(2)} \not= 0$ we have a P-solution
\item For $(\=M_m\LL\%\A\%\A)^{(2)} = 0$ and $\=M_m\LL\%\A\%\A \not= 0$ we have a Q-solution with antisymmetric dyadic $\=Q$
\item For  $\=M_m\LL\%\A\%\A = 0$ we have a P-solution with $\=P=P\=I$.
\end{itemize}

This result has the following conclusion. When the medium bidyadic $\=M_m$ is not restricted by a dispersion equation, i.e., it satisfies $(\=M_m\LL\%\nu\%\nu)^{(3)}=0$ for any one-form $\%\nu$, it has a Q-medium solution only when $\=Q$ is an antisymmetric dyadic, i.e., it belongs to case 1 solutions. In other words, there are no Q-medium solutions of full rank in case 2.

\end{document}